\documentclass[12pt,letterpaper]{article}
\pdfoutput=1
\usepackage{jheppub}
\usepackage{amsfonts, amsthm}
\usepackage[english]{babel}
\usepackage[utf8]{inputenc}
\usepackage{slashed}
\hypersetup{unicode}

\newcommand{\eq}{\begin{equation}}
\newcommand{\feq}{\end{equation}}
\newcommand{\eqn}{\begin{eqnarray}}
\newcommand{\feqn}{\end{eqnarray}}

\newcommand{\ma}[1]{\mbox{$\mathcal{#1}$}}

\title{Multi-centered black holes with a negative cosmological constant}

\author{Samuele Chimento}
\author{and Dietmar Klemm}
\affiliation{Dipartimento di Fisica, Universit\`a di Milano, and \\
INFN, Sezione di Milano, \\
Via Celoria 16, 20133 Milano, Italy.
}
\emailAdd{samuele.chimento@mi.infn.it}
\emailAdd{dietmar.klemm@mi.infn.it}
\preprint{IFUM-1020-FT}

\abstract{We present a recipe that allows to construct multi-centered black holes embedded in an
arbitrary FLRW universe. These solutions are completely determined by a function satisfying
the conformal Laplace equation on the spatial slices $\text{E}^3$, $\text{S}^3$ or $\text{H}^3$.
Since anti-de~Sitter space can be written in FLRW coordinates, this includes as a special case
multi-centered black holes in AdS, in the sense that, far away from the black holes, the energy
density and the pressure approach the values given by a negative cosmological constant.
We study in some detail the physical properties of the single-centered asymptotically AdS case,
which does not coincide with the usual Reissner-Nordstr\"om-AdS black hole, but is highly
dynamical. In particular, we determine the curvature singularities and trapping horizons of this
solution, compute the surface gravity of the trapping horizons, and show that the generalized first
law of black hole dynamics proposed by Hayward holds in this case. It turns out that the
spurious big bang/big crunch singularities that appear when one writes AdS in FLRW form,
become real in presence of these dynamical black holes. This implies that actually only one point
of the usual conformal boundary of AdS survives in the solutions that we construct.
Finally, a generalization to arbitary dimension is also presented.
}


\begin{document}
\maketitle
\flushbottom

\section{Introduction}

Composite objects formed by elementary constituents with mass to charge ratio equal to one
have been studied for a long time in general relativity, and more recently in supergravity
and string theory. While in the Newtonian theory of gravity it is clear that static
equilibrium for a system of point charges can be achieved by fine-tuning the charge suitably with the
particle mass, and thus balancing the mutual gravitational and electrostatic forces, the existence
of such static composite configurations is far from obvious in general relativity, whose equations
of motion are highly nonlinear, and therefore there is a priori no reason for a superposition principle
to hold. Yet, indications that such a general relativistic analogue exists first emerged when
Weyl \cite{Weyl:1917gp} obtained a particular class of static electromagnetic vacuum fields,
later generalized independently by Majumdar \cite{Majumdar:1947eu} and
Papapetrou \cite{Papapetrou:1947xx}\footnote{For higher-dimensional generalizations
see \cite{Lemos:2005md}.},
who removed Weyl's orginal restriction of axial symmetry.
In vacuum, the stationary generalization of the Majumdar-Papapetrou (MP) solution was constructed
by Israel, Wilson and Perj\'es \cite{Israel:1972vx,Perjes:1971gv}. In the same year, Hartle and
Hawking \cite{Hartle:1972ya} showed that the vacuum MP spacetime can describe a system of
multi-centered extremal Reissner-Nordstr\"om black holes.

These multi-black hole geometries admit supercovariantly constant spinors \cite{Gibbons:1982fy},
a result that actually extends to all the solutions belonging to the MP class \cite{Tod:1983pm}.
The MP solution can thus be seen as an early example of a BPS configuration that satisfies
rather simple first-order equations, and this explains also why one can build arbitrary superpositions
of the elementary constituents, in spite of the nonlinear nature of the Einstein-Maxwell equations.
Nevertheless, supersymmetry does not seem to be necessary for the existence of these bound states,
since by now we know many examples of multi-centered black holes that are not BPS,
cf.~e.g.~\cite{Bena:2009en}.

The study of composite systems like `black hole molecules' has played a crucial role in
several recent developments of supergravity and string theory, especially in attempts to
understand the quantum structure of black holes. Moreover, they are of interest in the field
of holography, in particular for applications of the gauge/gravity correspondence to condensed
matter phenomena (cf.~e.g.~\cite{Hartnoll:2009sz} for a review). In this context, it was established
recently in \cite{Anninos:2013mfa} that stable and metastable stationary bound states in
four-dimensional anti-de~Sitter space exist, and it was argued that their holographic duals represent
structural glasses. The glassy feature of these black hole bound states is related to their rugged free
energy landscape, which in turn is a consequence of the fact that the constituents can have a wide
range of different possible charges \cite{Anninos:2013mfa}.

Multi-centered black holes can also be generalized to dynamical situations.
Kastor and Traschen (KT) \cite{Kastor:1992nn} showed that the MP solution can be embedded in
a de~Sitter universe, where the no-force condition implies that the whole system is just
comoving with the cosmological expansion. The KT solution was then used in \cite{Brill:1993tm}
to study analytically black hole collisions. The embedding of composite black holes in
higher-dimensional de~Sitter spaces or in more general classes of FLRW universes was subject
of \cite{Klemm:2000vn,Klemm:2000gh,Maeda:2009zi,Maeda:2010ja,Gibbons:2009dr,
Chimento:2012mg}\footnote{Multi-black hole systems in Euclidean AdS were obtained
in \cite{Liu:2000ah}. However, these have no Lorentzian analogue.}.

Here we go one step further, and show how to construct multi-center solutions in any
FLRW spacetime and for arbitrary dimension. These geometries are sourced by a $\text{U}(1)$
gauge field and by a perfect fluid. Since anti-de~Sitter space can be written in an FLRW form (with
hyperbolic spatial slices and trigonometric scale factor), our recipe allows, as a particular subcase,
to obtain multi-center solutions in AdS. Like the underlying FLRW universe, these are highly
dynamical, and thus different in spirit from the bound states of \cite{Anninos:2013mfa}.
Generically, the $(D+1)$-dimensional black holes that we construct are determined by a function
satisfying the conformal Laplace equation on the spatial slices $\text{E}^D$, $\text{S}^D$ or $\text{H}^D$
of the FLRW background universe. This generalizes the well-known fact that asymptotically
flat extremal black holes are characterized by harmonic functions. Unfortunately, the spurious
big bang/big crunch singularities that appear when one writes AdS in FLRW coordinates,
become real once such a dynamical black hole is present. We show that this implies that actually
only one point of the conformal boundary of AdS survives. This makes it questionable
if our solutions admit an AdS/CFT interpretation in the usual sense.

The remainder of this paper is organized as follows: In the next section, starting from the
charged generalization of the McVittie spacetime \cite{McVittie:1933zz} found a long time ago
by Shah and Vaidya \cite{Vaidya:1968zza}, we show how to construct multi-center solutions in
an arbitrary FLRW universe. In section \ref{sing-hor-AdS}, we discuss some physical properties
of the single-centered nonextremal solution in AdS. In particular, we determine the curvature
singularities and trapping horizons, compute the surface gravity of the latter, and show that the
generalized first law of black hole dynamics proposed by Hayward \cite{Hayward:1997jp} holds.
In section \ref{higher-d}, the higher-dimensional case is considered, and in \ref{final-rem} we
present our conclusions.

\section{Multi-centered maximally charged McVittie solutions\label{sec:multi}}

In \cite{Vaidya:1968zza}, Shah and Vaidya presented a charged generalization of
the McVittie solution \cite{McVittie:1933zz}, with metric and $\text{U}(1)$ field strength given by
\begin{align}
 ds^2=&\frac{\left[1-(M^2-Q^2)\frac{1+kr^2}{4\, a^2\,r^2}\right]^2}{\left[1+M\frac{\sqrt{1+kr^2}}{a\,r}+(M^2-Q^2)\frac{1+kr^2}{4\, a^2\,r^2}\right]^2}dt^2\nonumber\\
      &-4\, a^2\left[1+M\frac{\sqrt{1+kr^2}}{a\,r}+(M^2-Q^2)\frac{1+kr^2}{4\, a^2\,r^2}\right]^2\frac{dr^2+r^2 d\theta^2+r^2\sin^2\theta d\phi^2}{(1+kr^2)^2}\,, \nonumber\\
 F=&\frac{Q}{a r^2}\frac{1}{\sqrt{1+kr^2}}\frac{\left[1-(M^2-Q^2)\frac{1+kr^2}{4\,a^2\,r^2}\right]}{\left[1+M\frac{\sqrt{1+kr^2}}{a\,r}+(M^2-Q^2)\frac{1+kr^2}{4\,a^2\,r^2}\right]^2}dr\wedge dt\,. \label{eq:nonmax_mcvittie}
\end{align}
\eqref{eq:nonmax_mcvittie} satisfies the Einstein-Maxwell equations
\eq
G_{\mu\nu} = 8\pi T_{\mu\nu}\,, \qquad \nabla_{\nu}F^{\mu\nu} = 4\pi J^{\mu}\,, \label{eq:Einst-Maxw}
\feq
\begin{displaymath}
T_{\mu\nu} = \frac1{4\pi}\left[-F_{\mu\rho}{F_{\nu}}^{\rho} + \frac14 g_{\mu\nu}F_{\rho\lambda}
F^{\rho\lambda}\right] + \rho u_{\mu}u_{\nu} + p(u_{\mu}u_{\nu} - g_{\mu\nu})\,, \qquad
J^{\mu} = \sigma u^{\mu}\,,
\end{displaymath}
where the pressure, energy density, charge density and four-velocity of the charged perfect fluid source
read respectively
\begin{align}
 8\pi p=&-2\left( \frac{\ddot a}{a} -\frac{\dot a^2}{a^2}\right)\frac{\left[1+M\frac{\sqrt{1+kr^2}}{a\,r}+(M^2-Q^2)\frac{1+kr^2}{4\,a^2\,r^2}\right]}{\left[1-(M^2-Q^2)\frac{1+kr^2}{4\,a^2\,r^2}\right]}-3\frac{\dot a^2}{a^2}\nonumber\\
 -k&\left\{a^2\left[1+M\frac{\sqrt{1+kr^2}}{a\,r}+(M^2-Q^2)\frac{1+kr^2}{4\,a^2\,r^2}\right]^2\left[1-(M^2-Q^2)\frac{1+kr^2}{4\,a^2\,r^2}\right]\right\}^{-1}\,, \nonumber
\end{align}
\begin{align}
 8\pi\rho=&3\frac{\dot a^2}{a^2} + \frac{3k}{2a^2}\left[1+M\frac{\sqrt{1+kr^2}}{a\,r}+(M^2-Q^2)\frac{1+kr^2}{4\,a^2\,r^2}\right]^{-3}\left[2+M\frac{\sqrt{1+kr^2}}{a\,r}\right]\,, \nonumber \\
 4\pi\sigma=&-\frac34 \frac{k Q}{a^3} \frac{\sqrt{1+kr^2}}{r}\left[1+M\frac{\sqrt{1+kr^2}}{a\,r}+(M^2-Q^2)\frac{1+kr^2}{4\, a^2\,r^2}\right]^{-3}\,, \label{eq:charge_dens}
\end{align}
\begin{equation}
 u=\frac{1-(M^2-Q^2)\frac{1+kr^2}{4\, a^2\,r^2}}{1+M\frac{\sqrt{1+kr^2}}{a\,r}+(M^2-Q^2)
      \frac{1+kr^2}{4\, a^2\,r^2}}dt\,.
\end{equation}
Moreover, $k=0,\pm 1$ determines the geometry of the spatial slices. From \eqref{eq:charge_dens} it is
clear that the cosmic fluid is required to be charged if the spatial geometry of the underlying
FLRW universe is curved.

In the maximally charged case $M=|Q|$ (obtained in \cite{Vaidya:1966zzc}), after the coordinate change
$r=\frac{1}{\sqrt{k}}\tan \frac{\sqrt{k}\,\psi}{2}$, \eqref{eq:nonmax_mcvittie} boils down to
\begin{align}
 ds^2=&\frac{1}{\left[1+M\frac{\sqrt{k}}{a\,\sin(\sqrt{k}\,\psi/2)}\right]^2}dt^2 \nonumber\\
      &-a^2\left[1+M\frac{\sqrt{k}}{a\,\sin(\sqrt{k}\,\psi/2)}\right]^2\left[d\psi^2+\frac{\sin^2(\sqrt{k}\psi)}{k}\left(d\theta^2+\sin^2\theta d\phi^2\right)\right]\,, \label{eq:max_mcvittie} \\
 F=&\frac{M k}{2a}\frac{\cos(\sqrt{k}\,\psi/2)}{\sin^2(\sqrt{k}\,\psi/2)}\frac{d\psi\wedge dt}{\left[1+
 M\frac{\sqrt{k}}{a\,\sin(\sqrt{k}\,\psi/2)}\right]^2} =
 d\left[\left(1+M\frac{\sqrt{k}}{a\,\sin(\sqrt{k}\,\psi/2)}\right)^{-1}dt\right]\,, \nonumber
\end{align}
while the pressure, energy- and current density become
\begin{align}
 8\pi p=&-3\frac{\dot a^2}{a^2}-\frac{k}{a^2}\left[1+M\frac{\sqrt{k}}{a\,\sin(\sqrt{k}\,\psi/2)}\right]^{-2}-2\left( \frac{\ddot a}{a} -\frac{\dot a^2}{a^2}\right)\left[1+M\frac{\sqrt{k}}{a\,\sin(\sqrt{k}\,\psi/2)}\right]\,, \nonumber\\
 8\pi\rho=&3\frac{\dot a^2}{a^2}+\frac32 \frac{k}{a^2}\left[1+M\frac{\sqrt{k}}{a\,\sin(\sqrt{k}\,\psi/2)}\right]^{-3}\left[2+M\frac{\sqrt{k}}{a\,\sin(\sqrt{k}\,\psi/2)}\right]\,, \nonumber\\
 4\pi J=&-\frac34 \frac{kM}{a^3}\frac{\sqrt{k}}{\sin(\sqrt{k}\,\psi/2)}\left[1+M\frac{\sqrt{k}}{a\,\sin(\sqrt{k}\,\psi/2)}\right]^{-4}dt\,.
\end{align}
This solution appears to be characterized by the function $H=\frac{M \sqrt{k}}{\sin(\sqrt{k}\,\psi/2)}$, which 
happens to satisfy the conformal Laplace equation on $\text{E}^3$, $\text{S}^3$ or $\text{H}^3$,
\begin{equation}
 \nabla^2 H = \frac18 R H\,,\label{eq:mcvittie_laplacian}
\end{equation}
where $R=6k$ is the corresponding scalar curvature. It is straightforward to verify that one can take
any function $\cal H$ solving \eqref{eq:mcvittie_laplacian}, and the resulting fields still satisfy the
Einstein-Maxwell equations \eqref{eq:Einst-Maxw}. This allows to generalize (\ref{eq:max_mcvittie}) to
a multi-centered solution by choosing $\cal H$ to be a linear combination of terms obtained by acting on
$H$ with the isometries of the three-dimensional base space metric.
Alternatively, one can use the conformal invariance of \eqref{eq:mcvittie_laplacian}, which implies
\eq
\tilde\nabla^2\tilde H = \frac18\tilde R\tilde H\,,
\feq
where $\tilde\nabla^2$ and $\tilde R$ denote the Laplacian and scalar curvature of the
conformally related metric $\tilde g_{ij}=\Omega^2 g_{ij}$ respectively, and
$\tilde H=\Omega^{-1/2}H$. Now let $g_{ij}$ be the flat metric, $g_{ij}dx^i dx^j=d\vec x^{\,2}$, and
\begin{displaymath}
\tilde g_{ij}dx^i dx^j = \frac{4d\vec x^{\,2}}{\left[1 + k\vec x^{\,2}\right]^2}\,.
\end{displaymath}
Starting from the usual one-center solution for a flat base, $H=\sqrt2 M/|\vec x|$, one gets
\begin{displaymath}
\tilde H = \frac{M}{|\vec x|}\sqrt{1 + k\vec x^{\,2}} = \frac{M\sqrt k}{\sin(\sqrt k\psi/2)}\,,
\end{displaymath}
which is the function appearing in \eqref{eq:max_mcvittie}. Taking instead
\begin{displaymath}
{\cal H} = \sum_{I=1}^N \frac{Q_I}{|\vec x - \vec x_I|}
\end{displaymath}
leads to
\eq
\tilde{\cal H} = \frac1{\sqrt 2}\left[1 + k\vec x^{\,2}\right]^{1/2}\sum_{I=1}^N\frac{Q_I}{|\vec x - \vec x_I|}\,.
\feq
It would be interesting to understand whether there is a deeper reason for the appearance
of this conformal structure.

Notice that the existence of this multi-centered generalization of \eqref{eq:max_mcvittie} is also
suggested by considering a charged probe particle in the geometry \eqref{eq:max_mcvittie},
whose equation of motion is
\eq
\nabla_v p^{\mu} = -q{F^{\mu}}_{\nu}v^{\nu}\,.
\feq
If the particle is BPS, $m=q$, and we take $v=v^t\partial_t$ for its four-velocity, it is easy to show
that the attractive gravitational force encoded in the Christoffel connection exactly cancels the
repulsive Lorentz force, such that the particle can stay at rest at fixed $\psi,\theta,\phi$.

\section{Singularities and horizons in the single-centered  asymptotically AdS case}
\label{sing-hor-AdS}

In this section, we shall discuss some physical properties of the single-centered (non
necessarily maximally charged) solution
in AdS, which does not coincide with the well-known Reissner-Nord\-str\"om-AdS black hole,
but is highly dynamical.

\begin{figure}[htb]
\begin{center}
\includegraphics[width=0.5\textwidth]{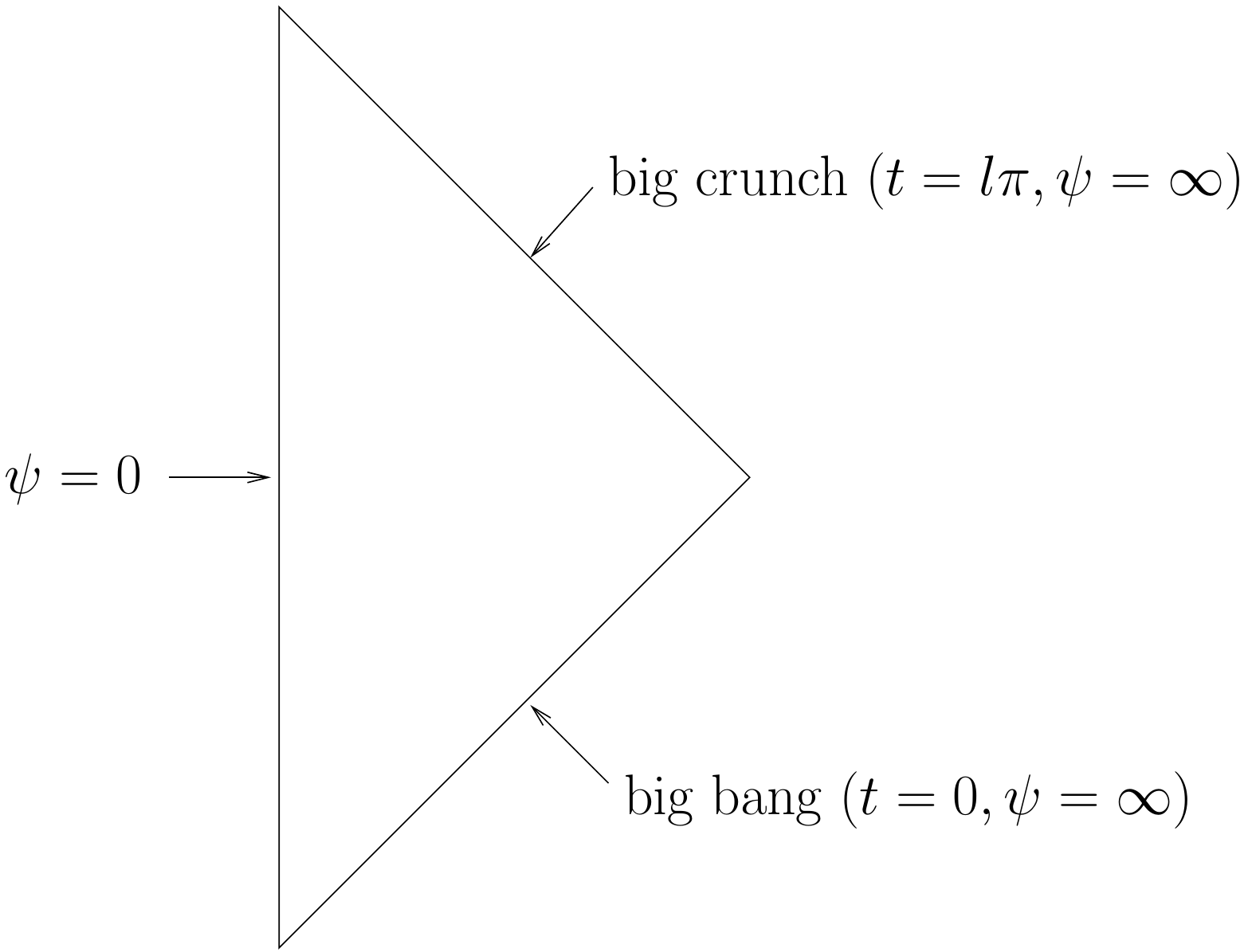}
\end{center}
\caption{Carter-Penrose diagram for AdS in FLRW coordinates.}
\label{Penrose-AdS-FLRW}
\end{figure}

Let us choose $k=-1$ and $a(t)=l\sin (t/l)$, with $l>0$ and $0<t/l<\pi$. Then, far from the black hole
($\psi\rightarrow\infty$ or $r\to 1$), the energy density and pressure approach the values given
by a negative cosmological constant $\Lambda=-3/l^2$, while the charge density \eqref{eq:charge_dens}
goes to zero. In this limit, the metric in \eqref{eq:nonmax_mcvittie} tends to AdS in FLRW coordinates, i.e.,
\eq
ds^2 \rightarrow dt^2 - l^2\sin^2\!\frac tl\left(d\psi^2 + \sinh^2\!\psi d\Omega^2\right)\,. \label{eq:AdS-FLRW}
\feq
The FLRW form is related to global coordinates $\tau, \hat r$ by
\eq
\hat r = l\sin\frac tl\sinh\psi\,, \qquad \cos\frac tl = \left(1 + \frac{\hat r^2}{l^2}\right)^{1/2}\!\cos\frac{\tau}{l}\,,
             \label{eq:FLRW-global}
\feq
which casts \eqref{eq:AdS-FLRW} into
\eq
ds^2 = \left(1 + \frac{\hat r^2}{l^2}\right)d\tau^2 - \left(1 + \frac{\hat r^2}{l^2}\right)^{-1}d\hat r^2
            -  \hat r^2 d\Omega^2\,.
\feq
\eqref{eq:AdS-FLRW} has a lightlike big bang/big crunch singularity in $t=0$ and $t=l\pi$ respectively,
that are of course artefacts of the coordinate system $t,\psi$. In fact, by introducing $\tau,\hat r$,
one extends the spacetime beyond these singularities. The causal structure of AdS in FLRW
coordinates is visualized in the Carter-Penrose diagram fig.~\ref{Penrose-AdS-FLRW}.

Notice also that, due to $\cos^2(t/l)\le 1$, the last eq.~of \eqref{eq:FLRW-global} implies
$\tau/l\rightarrow\pi/2$ for $\hat r\to\infty$, so that actually only the point $\tau=l\pi/2$
(which is of course a two-sphere) of the conformal boundary of AdS is visible in FLRW
coordinates.

Rewriting the metric (\ref{eq:nonmax_mcvittie}) for 
brevity as
\begin{equation}
 ds^2=\frac{g^2}{f^2}dt^2 - a^2 f^2\left( d\psi^2+\frac{\sin^2(\sqrt{k}\psi)}{k} d\Omega^2
\right)\,, \nonumber\\
\end{equation}
with 
\begin{equation}
 f=1+\frac{\sqrt{k}M}{a\sin(\sqrt{k}\psi/2)}+k\frac{M^2-Q^2}{4\, a^2\sin^2(\sqrt{k}\psi/2) },\qquad g=1-k\frac{M^2-Q^2}{4\, a^2\sin^2(\sqrt{k}\psi/2) }\,,
\end{equation}
the scalar curvature is
\begin{equation}
 R=-12 \frac{\dot a^2}{a^2}-6\frac{f}{g} \left(\frac{\ddot a}{a}-\frac{\dot a^2}{a^2} \right)-\frac{3}{2}k\frac{f(g+2)+g^2}{a^2 f^3 g}\,.
\end{equation}
The spacetime with $k=-1$ has thus curvature singularities in $a(t)=0$, 
$\sinh(\psi/2)=\pm \frac{\sqrt{M^2-Q^2}}{2 a}$ and $\sinh(\psi/2)=\frac{\pm Q-M}{2 a}$; however the only singularity 
that is connected with the asymptotic region $\psi\rightarrow +\infty$ is the hypersurface 
$\sinh(\psi/2)=\frac{\sqrt{M^2-Q^2}}{2 a}$. In the maximally charged case, $M=|Q|$, this singular hypersurface becomes the 
union of the hypersurfaces $t=0$, $t=l\pi$ and $\psi=0$.

To determine if the present spacetime describes a black hole, one can look for trapping horizons \cite{Hayward:1993wb}.
Introducing the Newman-Penrose null tetrads 
\begin{align}
 l=&\frac{1}{\sqrt{2}} \left( \frac{g}{f} dt-a f d\psi \right)\,, \qquad
 n=\frac{1}{\sqrt{2}} \left( \frac{g}{f} dt+a f d\psi \right)\,, \nonumber \\
 m=&\frac{a f \sinh\psi}{\sqrt{2}}\left( d\theta+i\sin\theta d\varphi \right)\,,
\end{align}
and the complex conjugate $\bar m$, the expansions of the outgoing and ingoing radial null
geodesics are respectively
\begin{equation}
  \theta_+\equiv-2m^{(\mu}\bar m^{\nu)}\nabla_\mu l_\nu\,,\qquad \theta_-\equiv-2m^{(\mu}\bar m^{\nu)}\nabla_\mu n_\nu\,,
\end{equation}
and once evaluated read
\begin{equation}
 \theta_\pm=\frac{\sqrt{2}}{a}\left[\dot a\pm\frac{g+\sinh^2(\psi/2)(f+g)}{\sinh\psi \,f^2}\right].\label{eq:expansions}
\end{equation}
Marginal surfaces are defined as spacelike 2-surfaces on which $\theta_+=0$ ($\theta_-=0$), and trapping horizons are 
defined as the closure of 3-surfaces foliated by marginal surfaces such that $\theta_-\neq 0$ and 
$\ma{L}_-\theta_+\neq 0$ ($\theta_+\neq 0$ and $\ma{L}_+\theta_-\neq 0$) on the 3-surface, where $\ma{L}_\pm$ is the Lie
derivative along the outgoing or ingoing radial null geodesics.
From eq.~(\ref{eq:expansions}) it is clear that if $t\neq l\, \pi/2$ the two expansions can't both vanish at the same time,
while in $t= l \,\pi/2$ they only vanish behind or on the singularity, since outside of the singularity both $f$ and $g$ 
are positive, so that no horizon can exist in any case for $t= l \,\pi/2$. Furthermore $\ma{L}_-\theta_+$ and 
$\ma{L}_+\theta_-$ are negative in the whole considered region; as a consequence the only condition necessary to locate 
the trapping horizons is the vanishing of $\theta_+$ or $\theta_-$.

For $M\neq|Q|$ there are always solutions to $\theta_\pm =0$ that lie on the singularity; this means
that the horizons intersect the singularity and there is a time interval around $t= l \,\pi/2$ for which
they are not defined. On the other hand, if $M=|Q|$ the horizons are defined for every $t\neq l \,\pi/2$,
while for $t= l \,\pi/2$ they tend to coincide on the 
singularity $\psi=0$. For $\psi\rightarrow+\infty$, $\theta_\pm =0$ implies $\dot a\rightarrow \pm 1$ which means that
the horizons tend to the axes $t=0$ and $t= l\,\pi$.

There are always two trapping horizons: One for $t>l \,\pi/2$ where $\theta_+=0$ and 
$\theta_-=2\sqrt{2}\frac{\dot a}{a}<0$, and the other for $t<l \,\pi/2$ where $\theta_-=0$ and 
$\theta_+=2\sqrt{2}\frac{\dot a}{a}>0$. Since $\ma{L}_-\theta_+$ and $\ma{L}_+\theta_-$ are
negative these are respectively an outer future trapping horizon, which can be interpreted as the horizon of a black hole,
and an outer past trapping horizon, which can be interpreted as the horizon of a white hole.

\begin{figure}[htb]
 \begin{center}
  \includegraphics[scale=0.7]{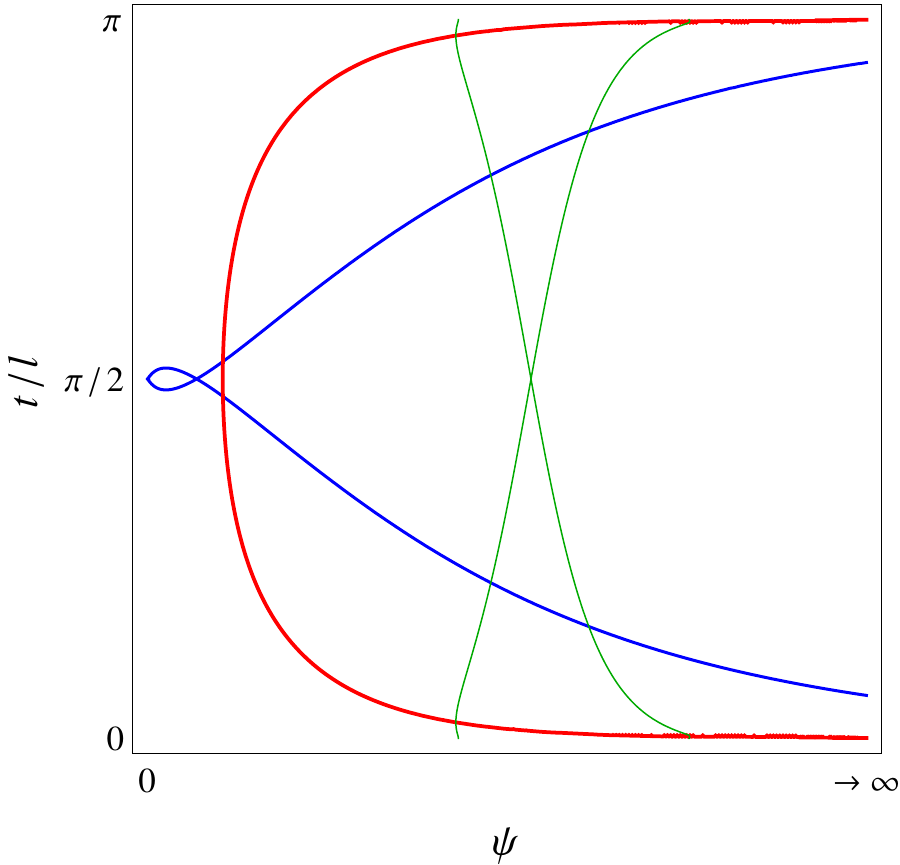}\hspace{1.5cm}\includegraphics[scale=0.7]{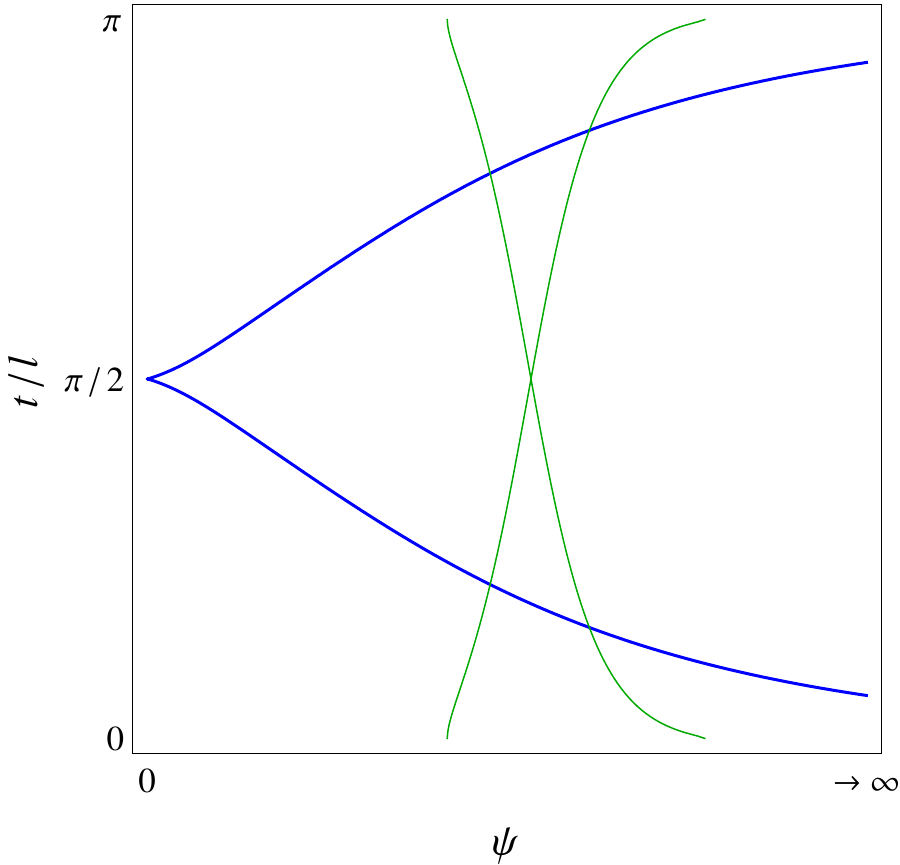}
 \end{center}
 \caption{Plots of curvature singularity (red), trapping horizons (blue) and one pair of radial null
geodesics (green) crossing in $t=l\pi/2$, in FLRW coordinates $(t,\psi)$ for $M\neq |Q|$ (left) and
$M=|Q|$ (right). For $M=|Q|$ the curvature singularities coincide with the axes $\psi=0$,
$t=0$ and $t=l\pi$. \label{horizons_psi}}
\end{figure}

\begin{figure}[htb]
 \begin{center}
  \includegraphics[scale=0.7]{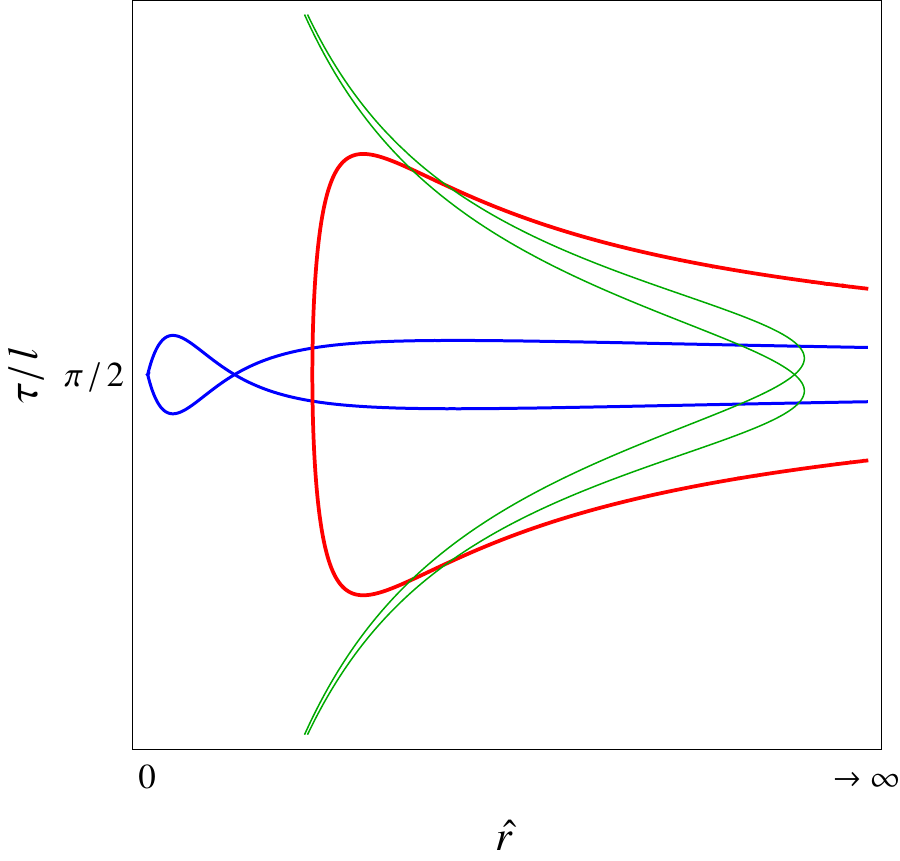}\hspace{1.5cm}\includegraphics[scale=0.7]{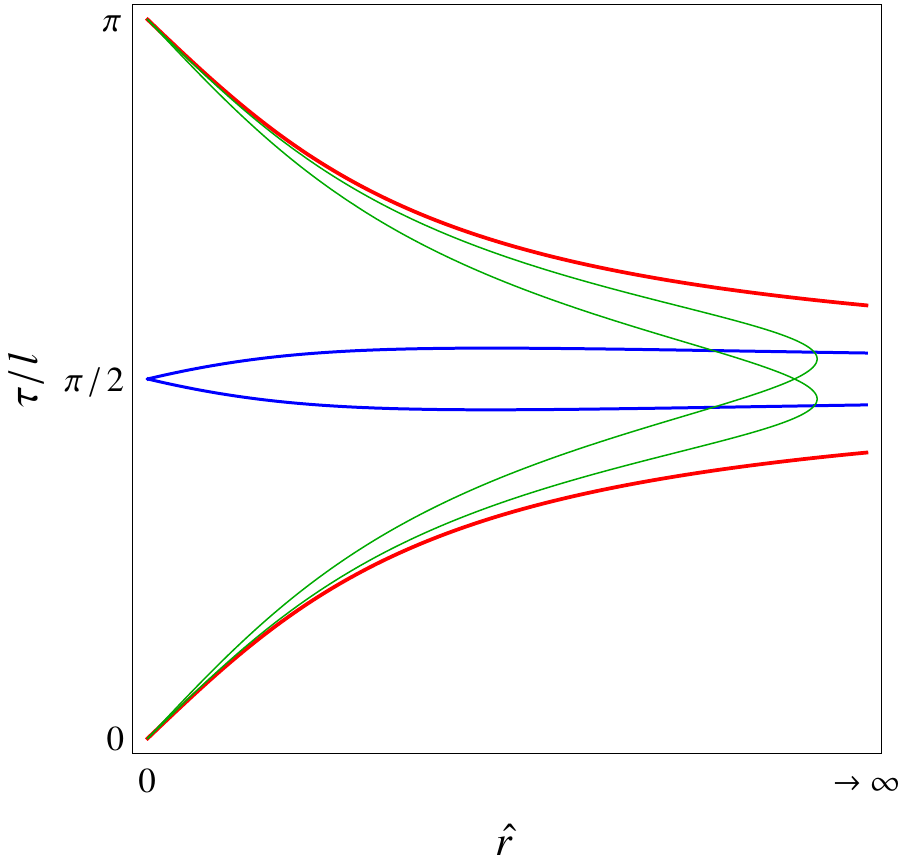}
 \end{center}
 \caption{Plots of curvature singularity (red), trapping horizons (blue) and one pair of radial null
geodesics (green) crossing in $\tau= l\pi/2$, in the coordinate system $(\tau,\hat r)$ for $M\neq |Q|$
(left) and $M=|Q|$ (right). The plot for $M\neq |Q|$ is zoomed in on the vertical axis to show its relevant features. For $M=|Q|$, the axis $\hat r=0$ belongs to the curvature singularity. \label{horizons_glob}}
\end{figure}

In figures \ref{horizons_psi} and \ref{horizons_glob} we display, respectively in the cosmological (FLRW)
coordinates $(t,\psi)$ and in the global coordinates $(\tau,\hat r)$ as defined in (\ref{eq:FLRW-global}), the
curvature singularity, the trapping horizons and the radial null geodesics intersecting in a point
with $t=l\pi/2$ or $\tau=l\pi/2$, for arbitrarily chosen parameters; the plots are obtained by
numerical methods.

The radial null geodesics satisfy
\begin{equation}
 \frac{dt}{d\psi}=\pm \frac{a f^2}{g}\,.
\end{equation}
For the case $M=|Q|$ this means that the singularity at $\psi=0$ is never reached, since for finite $a$ the derivative 
tends to infinity. On the other hand the $t$-component of the geodesic equation for radial null or timelike geodesics in 
$a\sim 0$ and finite $\psi$, using $|\dot \psi|\le |\dot t|/(a f^2)$ reads
\begin{equation}
 \ddot t + \frac{\cos(t/l)}{l\sin(t/l)} \dot t^2\sim 0\,,
\end{equation}
where a dot indicates a derivative with respect to the affine parameter. The solution, 
$t\sim\pm\, l \cos^{-1}\left(c_1 \lambda + c_2 \right)$ shows that the singularities in $t=0,\,l\pi$
are always reached for finite values of the affine parameter.

Taking advantage of the spherical symmetry, it is possible to define in a simple, geometrical way the
surface gravity $k_l$ on the trapping horizons \cite{Hayward:1997jp} and the associated local Hawking
temperature $T=\frac{k_l}{2\pi}$ \cite{Hayward:2008jq}, according to
\begin{gather}
 k_l=-\frac{1}{2}\left.\tilde\nabla_\mu\tilde\nabla^\mu \ma{R}\right|_{\theta_\pm=0}=-\frac{\ma{R}}{2} \left[\frac{f}{g} \left( \frac{\ddot a}{a}-\frac{\dot a^2}{a^2} \right)+2\frac{\dot a^2}{a^2}\right.\nonumber\\
 \left.+\frac12\frac{1}{a^2 f^3 \sinh^2(\psi/2)}\left( \frac32 g-\cosh^2(\psi/2) \right) \pm\frac{\dot a}{a^2 f}\frac{\cosh(\psi/2)}{\sinh(\psi/2)}\left( \frac12+\frac1g-\frac{g}{f} \right)\right]\,,
\end{gather}
where $\ma{R}=a f \sinh\psi$ is the areal radius, $\tilde\nabla$ is the covariant derivative operator associated with the 
two-dimensional metric normal to the spheres of symmetry, and the vanishing of expression (\ref{eq:expansions}) was used. $k_l$ is in general not zero even in the maximally charged case, and
is positive on the horizons, as is expected for outer trapping horizons. It is straightforward to verify that
the generalized first law of black hole dynamics proposed by Hayward in \cite{Hayward:1997jp},
\begin{equation}
 E'=\frac{k_l A'}{8\pi}+\frac12\ma{T} V'\,,
\end{equation}
holds on the trapping horizons. Here a prime represents a derivative along a vector field tangent to the
trapping horizon, $A=4\pi\ma{R}^2$ is the area of the spheres of symmetry, $V=\frac43\pi\ma{R}^3$
is the areal volume, $\ma{T}$ is the trace of the total energy-momentum tensor $T$ with respect to the
two-dimensional normal metric, and $E$ is the Misner-Sharp energy, defined as
\begin{equation}
 E=\frac{1}{2}\ma{R}\left( 1+\nabla_\mu\ma{R}\nabla^\mu\ma{R} \right)\,.
\end{equation}
Notice that $\nabla_\mu\ma{R}\nabla^\mu\ma{R}=\theta_+\theta_-\,\ma{R}^2/2$ is identically zero
on the trapping horizons, implying $E'=\frac{1}{2}\ma{R}'$.

\section{Higher-dimensional generalization}
\label{higher-d}

It is possible to construct higher-dimensional generalizations of the multi-centered solutions found in
section \ref{sec:multi}. To this aim, inspired by previous results \cite{Patel:1999ej, Gao:2004gz},
we use the ansatz
\begin{eqnarray}
 ds^2 &=& \frac{g^2}{f^2}dt^2 - a^2 f^{\frac{2}{D-2}}ds_D^2\,, \label{eq:hd_nonext} \\
F &=& \sqrt{\frac{D-1}{2(D-2)}}\frac{g}{f}\left[\left( 1-\frac{g}{f} \right)^2+4\frac{g}{f}\left( 1-
\frac1g \right)\right]^{1/2}\frac{dH}{H}\wedge dt\,, \nonumber
\end{eqnarray}
with
\begin{equation}
 f=1+M\frac{H}{a^{D-2}}+\frac{M^2-Q^2}{4}\frac{H^2}{a^{2(D-2)}}\,,\qquad g=1-\frac{M^2-Q^2}{4}\frac{H^2}{a^{2(D-2)}}\,,
\end{equation}
where $a(t)$ is a function of time, $H(\vec x)$ is a function of the spatial coordinates, $D$ and 
$ds_D^2\equiv h_{ij}dx^i dx^j$ are respectively the dimension and the metric of the spatial slices.
Notice that the square bracket in the expression of $F$ is equal to $Q H/(a^{D-2}f)$ and is just
a way to express the charge $Q$ in terms of the functions $f$ and $g$. 

The nonvanishing components of the Einstein tensor for \eqref{eq:hd_nonext} are given by
\begin{align}
 G_{tt}=&\frac{D(D-1)}{2}\frac{\dot a^2}{a^2}\frac{g^2}{f^2}+\frac{\hat R}{2 a^2}\frac{g^2}{f^{2\frac{D-1}{D-2}}}-\frac{D-1}{D-2}\left(1-\frac{g}{f}\right)\frac{g^2}{f^{2\frac{D-1}{D-2}}}\frac{\hat\nabla^2 H}{a^2 H}\nonumber\\
 +\left\{\vphantom{\left(1-\frac{g}{f}  \right)^2}\right.&\left.\mkern-7mu\frac{2}{D-2}\frac{1}{f}\left(1-\frac{1}{g}  \right)\left[ g(D-1)+D-3 \right]+\frac{1}{2}\frac{D-1}{D-2}\left(1-\frac{g}{f}  \right)^2  \right\}\frac{g^2}{f^{2\frac{D-1}{D-2}}}\frac{\partial_l H h^{lm}\partial_m H}{a^2 H^2}\,, \nonumber\\
 G_{ij}=&(1-D)h_{ij}a^2\frac{f^\frac{D}{D-2}}{g}\left( \frac{\ddot a}{a} -\frac{\dot a^2}{a^2}\right)-\frac{D(D-1)}{2}\dot a^2f^\frac{2}{D-2}h_{ij}\nonumber\\
 &+\hat R_{ij}-\frac{1}{2}\hat R h_{ij}+2\left( 1-\frac{1}{g} \right)\left( \frac{\hat\nabla^2H}{H}h_{ij}-\frac{\hat\nabla_i\hat\nabla_jH}{H} \right)\nonumber\\
 &+\left\{ \frac{D-1}{2}\left( 1-\frac{g}{f} \right)^2-2\left( 1-\frac1g \right)\left[ 1-(D-1)\frac{g}{f} \right] \right\} \frac{1}{D-2}\frac{\partial_l H h^{lm}\partial_m H}{H^2}h_{ij}\nonumber\\
 &-\left\{ (D-1)\left( 1-\frac{g}{f} \right)^2-2\left( 1-\frac{1}{g} \right)\left[ D+2(1-D)\frac{g}{f} \right] \right\}\frac{1}{D-2}\frac{\partial_iH\partial_jH}{H^2}\,,
\end{align}
where $\hat\nabla$, $\hat R_{ij}$ and $\hat R$ represent respectively the covariant derivative,
Ricci tensor and scalar curvature of the spatial metric $h_{ij}$.
From the expression for $F$ one obtains for the electromagnetic energy-momentum tensor
\begin{align}
 8\pi T^{\text{em}}_{tt}&=\frac12\frac{D-1}{D-2}\left[ \left( 1-\frac{g}{f} \right)^2+4\frac{g}{f}\left( 1-\frac{1}{g} \right)\right]\frac{g^2}{f^{2\frac{D-1}{D-2}}}\frac{\partial_l H h^{lm}\partial_m H}{a^2 H^2}\,,\\
 8\pi T^{\text{em}}_{ij}&=-\frac{D-1}{D-2}\left[ \left( 1-\frac{g}{f} \right)^2+4\frac{g}{f}\left( 1-\frac{1}{g} \right)\right]\frac{1}{H^2}\left( \partial_iH\partial_jH-\frac{1}{2}h_{ij}\partial_l H h^{lm}\partial_m H \right)\,.
\nonumber
\end{align}
The requirement to have a perfect fluid as matter source translates into the condition
$G_{ij}-8\pi T^{\text{em}}_{ij}\propto h_{ij}$. This implies that $\hat R_{ij}\propto h_{ij}$, that is, the
spatial slices must be Einstein manifolds, and that the function $H$ must satisfy the condition
\begin{equation}
 -\frac{\hat\nabla_i\hat\nabla_jH}{H}+\frac{D}{D-2}\frac{\partial_iH\partial_jH}{H^2}\propto h_{ij}\,.\label{eq:H_cond}
\end{equation}
Notice that (\ref{eq:H_cond}) is conformally invariant \emph{on Einstein manifolds}, in the sense
that under a conformal transformation that maps $h_{ij}$ to $\tilde h_{ij}=e^{2\omega}h_{ij}$, assuming
that $H$ tranforms as $\tilde H=e^{\frac{2-D}{2}\omega} H$, one has
\begin{equation}
 -\frac{\tilde{\hat\nabla}_i\tilde{\hat\nabla}_j\tilde H}{\tilde H}+\frac{D}{D-2}\frac{\partial_i\tilde H\partial_j\tilde H}{\tilde H^2}=-\frac{\hat\nabla_i\hat\nabla_jH}{H}+\frac{D}{D-2}\frac{\partial_iH\partial_jH}{H^2}+\frac{\hat R_{ij}-\tilde{\hat R}_{ij}}{2}\,.
\end{equation}

For a metric, $\text{U}(1)$ gauge field, fluid velocity and current density of the form
\begin{equation}
ds^2 = V(t,x^i)dt^2 - g_{ij}dx^i dx^j\,, \qquad A = \phi dt\,, \qquad u = \sqrt V dt\,,
\qquad J = \rho_{\text e}dt\,,
\end{equation}
(which is precisely what we have here), the conservation laws $\nabla_{\mu}T^{\mu\nu}=0$
imply
\begin{equation}
\partial_t p + \frac{p+\rho}2 g^{ij}\partial_t g_{ij} = 0\,, \qquad \partial_i p + \frac{p+\rho}{2V}
\partial_i V - \frac{\rho_{\text e}}{\sqrt V}\partial_i\phi = 0\,.
\end{equation}
These equations carry information on how the pressure gradients balance the equilibrium of the
system. In particular, the second one shows that the spatial gradient of the pressure cancels the
gravitational and electromagnetic forces. Note that, due to the explicit time-dependence, there
is one additional equation w.r.t.~(17) of \cite{Lemos:2005md}.

Let us now turn to \eqref{eq:H_cond}.
In the particular case of a conformally flat spatial metric, $h_{ij}=e^{2\omega}\delta_{ij}$, for it to be
Einstein it must also be of constant curvature, and one can always take $e^{-\omega}=1+\frac{k}{4}r^2$,
with $r^2\equiv\sum x^ix^i$.
Then we have $H=\left(1+\frac{k}{4}r^2\right)^{\frac{D-2}{2}} H_0$, with $H_0$ satisfying
(\ref{eq:H_cond}) on flat space, i.e.,
\begin{equation}
 H_0=(\alpha r^2 + \beta^i x^i +\gamma)^\frac{2-D}{2}\,,
\end{equation}
where $\alpha$ can always be set to $1$ by rescaling the parameters $M$  and $Q$.
In this case the energy density and pressure of the fluid are given by
\begin{align}
 8\pi\rho=&\frac{f^2}{g^2}\left( G_{tt}-8\pi T^{\text{em}}_{tt} \right)=\frac{D(D-1)}{2}\frac{\dot a^2}{a^2}+\frac{k D (D-1)}{2 a^2}\frac{1}{f^\frac{2}{D-2}}\nonumber\\
 &-\frac{D-1}{D-2}\left(1-\frac{g}{f}\right)\frac{1}{f^\frac{2}{D-2}}\frac{\hat\nabla^2 H}{a^2 H}+2\frac{D-3}{D-2}\left( 1-\frac{1}{g} \right)\frac{1}{f^\frac{D}{D-2}}\frac{\partial_l H h^{lm}\partial_m H}{a^2 H^2}\,,\nonumber\\
 8\pi p=&\frac{h^{ij}}{D a^2 f^\frac{2}{D-2}}\left( G_{ij}-8\pi T^{\text{em}}_{ij} \right)=-\frac{k}{2 a^2}\frac{(D-1)(D-2)}{f^\frac{2}{D-2}}+(1-D)\frac{f}{g}\left( \frac{\ddot a}{a} -\frac{\dot a^2}{a^2}\right)\nonumber\\
 &-\frac{D(D-1)}{2}\frac{\dot a^2}{a^2}+\frac{2}{a^2 f^\frac{2}{D-2}}\left( 1-\frac{1}{g} \right)\frac{D-1}{D}\frac{\hat\nabla^2 H}{H}\,,
\end{align}
while the current density reads
\begin{gather}
 4\pi J=-\sqrt{\frac{D-1}{2(D-2)}}\left[\left( 1-\frac{g}{f} \right)^2+4\frac{g}{f}\left( 1-\frac1g \right)\right]^{1/2}\frac{g}{a^2 f^\frac{D}{D-2}}\frac{\hat\nabla^2 H}{H} dt\,.
\end{gather}

In the maximally charged case, $|Q|=M$, the ansatz (\ref{eq:hd_nonext}) reduces to
\begin{equation}
 ds^2=\frac{1}{f^2}dt^2 - a^2 f^{\frac{2}{D-2}}ds_D^2\,,\quad F=\sqrt{\frac{D-1}{2(D-2)}}\frac{1}{f}\left( 1-\frac{1}{f} \right)\frac{dH}{H}\wedge dt\,, \nonumber\\
\end{equation}
with
\begin{equation}
 f=1+M\frac{H}{a^{D-2}}\,,
\end{equation}
% with Ricci
% \begin{gather}
%  R_{tt}=-D\frac{\dot a^2}{a^2}\frac{1}{f^2}-D\left( \frac{\ddot a}{a} -\frac{\dot a^2}{a^2}\right)\frac{1}{f}-\frac{1-\frac1f}{f^{2\frac{D-1}{D-2}}}\left[\frac{\hat\nabla^2 H}{a^2 H}-\left( 1-\frac{1}{f} \right)\frac{\partial_l H h^{lm}\partial_m H}{a^2 H^2}\right]\nonumber\\
%  R_{ij}=\hat R_{ij}+a^2 h_{ij}f^\frac{D}{D-2}\left( \frac{\ddot a}{a} -\frac{\dot a^2}{a^2}\right)+D \dot a^2 h_{ij}f^\frac{2}{D-2}-\frac{h_{ij}}{D-2}\left( 1-\frac{1}{f}\right)\frac{\hat\nabla^2 H}{H}\nonumber\\
%  +\frac{1}{D-2}\left(1-\frac1f\right)^2\frac{\partial_l H h^{lm}\partial_m H}{H^2}h_{ij}-\frac{D-1}{D-2}\left( 1-\frac{1}{f} \right)^2\frac{\partial_i H\partial_j H}{H^2}\,,
% \end{gather}
% scalar curvature
% \begin{gather}
%  R=-D(D+1)\frac{\dot a^2}{a^2}-2D f\left( \frac{\ddot a}{a} -\frac{\dot a^2}{a^2}\right)+\frac{2}{D-2}\left(1-\frac{1}{f}  \right)\frac{1}{f^\frac{2}{D-2}}\frac{\hat\nabla^2 H}{a^2 H}\nonumber\\
%  +\left(1-\frac{1}{f}  \right)^2\frac{1}{f^\frac{2}{D-2}}\frac{\partial_l H h^{lm}\partial_m H}{a^2 H^2}\frac{D-3}{D-2}-\frac{\hat R}{a^2 f^\frac{2}{D-2}}\,,
% \end{gather}
and the Einstein tensor boils down to
\begin{align}
 G_{tt}=&\frac{D(D-1)}{2}\frac{\dot a^2}{a^2}\frac{1}{f^2}+\frac{\hat R}{2 a^2}\frac{1}{f^{2\frac{D-1}{D-2}}}-\frac{D-1}{D-2}\left(1-\frac{1}{f}\right)\frac{1}{f^{2\frac{D-1}{D-2}}}\frac{\hat\nabla^2 H}{a^2 H}\nonumber\\
 &+\frac{1}{2}\frac{D-1}{D-2}\left(1-\frac{1}{f}  \right)^2\frac{1}{f^{2\frac{D-1}{D-2}}}\frac{\partial_l H h^{lm}\partial_m H}{a^2 H^2}\,, \nonumber\\
 G_{ij}=&(1-D)h_{ij}a^2 f^\frac{D}{D-2}\left( \frac{\ddot a}{a} -\frac{\dot a^2}{a^2}\right)-\frac{D(D-1)}{2}\dot a^2f^\frac{2}{D-2}h_{ij}\nonumber\\
 &+\hat R_{ij}-\frac{1}{2}\hat R h_{ij}+\frac{D-1}{D-2}\left( 1-\frac{1}{f} \right)^2\left[\frac{1}{2}\frac{\partial_l H h^{lm}\partial_m H}{H^2}h_{ij}-\frac{\partial_iH\partial_jH}{H^2}\right]\,.
\end{align}
Finally, the electromagnetic energy-momentum tensor becomes
\begin{align}
 8\pi T^{\text{em}}_{tt}&=\frac12\frac{D-1}{D-2}\left( 1-\frac{1}{f} \right)^2\frac{1}{f^{2\frac{D-1}{D-2}}}\frac{\partial_l H h^{lm}\partial_m H}{a^2 H^2}\,, \nonumber\\
 8\pi T^{\text{em}}_{ij}&=-\frac{D-1}{D-2}\left( 1-\frac{1}{f} \right)^2\frac{1}{H^2}\left( \partial_iH\partial_jH-\frac{1}{2}h_{ij}\partial_l H h^{lm}\partial_m H \right)\,.
\end{align}
In this case the condition to have a perfect fluid source, $G_{ij}-8\pi T^{\text{em}}_{ij}\propto h_{ij}$,
simply reduces to the requirement that the spatial slices are Einstein manifolds, $\hat R_{ij}\propto h_{ij}$,
while $H$ can now be any function of the spatial coordinates, i.e., \eqref{eq:H_cond} does not need to
hold anymore. The maximally charged solution is thus less constrained.
This is of course also true in the four-dimensional case, with suitable forms for the
density, pressure and current of the fluid.
For a spatial metric of constant curvature the energy density, pressure and current density of the fluid
are respectively given by
\begin{align}
 8\pi\rho&=\frac{D(D-1)}{2}\frac{\dot a^2}{a^2}+\frac{k D (D-1)}{2 a^2}\frac{1}{f^\frac{2}{D-2}}-\frac{D-1}{D-2}\left(1-\frac{1}{f}\right)\frac{1}{f^\frac{2}{D-2}}\frac{\hat\nabla^2 H}{a^2 H}\,, \nonumber\\
 8\pi p&=-\frac{k}{2 a^2}\frac{(D-1)(D-2)}{f^\frac{2}{D-2}}+(1-D)f\left( \frac{\ddot a}{a} -\frac{\dot a^2}{a^2}\right)-\frac{D(D-1)}{2}\frac{\dot a^2}{a^2}\,, \nonumber \\
 4\pi J&=-\sqrt{\frac{D-1}{2(D-2)}}\left( 1-\frac{1}{f} \right)\frac{1}{a^2 f^\frac{D}{D-2}}\frac{\hat\nabla^2 H}{H} dt\,.
\end{align}
Given that $H$ can be an arbitrary function in the extremal case, what was the reason for the
appearance of the conformal Laplace equation in section \ref{sec:multi}? To answer this question,
let us go back to the nonextremal solution, and consider the case where $h_{ij}$ is the metric
on a space of constant curvature. As we already said, one has then (setting $\alpha=1$)
\begin{equation}
H = \left(1 + \frac k4 r^2\right)^{\frac{D-2}2}H_0\,, \qquad H_0 = (r^2 + \beta^i x^i +
\gamma)^{\frac{2-D}2}\,. \label{H-nonextr}
\end{equation}
If the parameters in \eqref{H-nonextr} satisfy the constraint $\gamma=\beta^i\beta^i/4$,
$H_0$ can be rewritten as
\begin{equation}
H_0 = \frac1{|\vec x - \vec x_0|^{D-2}}\,, \qquad (x^i_0 \equiv -\beta^i/2)\,,
\end{equation}
which is harmonic on $D$-dimensional flat space. In this case, $H$ in \eqref{H-nonextr}
satisfies the conformal Laplace equation
\begin{equation}
\hat\nabla^2 H = \frac{D-2}{4(D-1)}\hat R H\,. \label{conf-Laplace-D}
\end{equation}
\eqref{conf-Laplace-D} results thus from extrapolating the nonextremal case
(where \eqref{eq:H_cond} must hold) to the maximally charged situation, under the
additional assumption that $h_{ij}$ has constant curvature.

\section{Final remarks}
\label{final-rem}

In this paper, we showed how to construct multi-center black hole bound states in an
arbitrary FLRW universe, and for any dimension. It turned out that these
solutions are characterized by a function satisfying the conformal Laplace equation
on the spatial slices of the FLRW background. For the single-center solution, we discussed some
of the physical properties in the case when the energy density and the pressure of the perfect fluid
source approach the values given by a negative cosmological constant far away from the black hole.

It would be nice to mimic the perfect fluid with one or more scalar fields, and to
embed \eqref{eq:max_mcvittie} in some simple model of matter-coupled (genuine or fake)
$N=2$ supergravity, similar to what was done in
\cite{Maeda:2009zi,Maeda:2010ja,Gibbons:2009dr,Chimento:2012mg}\footnote{Note that, in
\cite{Maeda:2009zi,Maeda:2010ja,Gibbons:2009dr,Chimento:2012mg}, the stress tensor of the
scalar did not assume exactly a perfect fluid form everywhere, but only far away from the black holes.}.
Since the charge density $\sigma$ of the cosmic fluid is nonvanishing for
$k\neq 0$, these scalars have to be charged under a $\text{U}(1)$ gauge field. In such a scenario,
the cosmological expansion would be driven by the scalar field while rolling down its
potential.

Our results represent another example for a superposition principle without supersymmetry:
Generically, the existence of a Killing spinor implies a timelike or null Killing vector, which we
clearly don't have here.


\begin{thebibliography}{99}

%\cite{Weyl:1917gp}
\bibitem{Weyl:1917gp}
  H.~Weyl,
  ``The theory of gravitation,''
  Annalen Phys.\  {\bf 54} (1917) 117.
  %%CITATION = ANPYA,54,117;%%

%\cite{Majumdar:1947eu}
\bibitem{Majumdar:1947eu}
  S.~D.~Majumdar,
  ``A class of exact solutions of Einstein's field equations,''
  Phys.\ Rev.\  {\bf 72} (1947) 390.
  %%CITATION = PHRVA,72,390;%%

\bibitem{Papapetrou:1947xx}
  A.~Papapetrou,
  ``A static solution of the equations of the gravitational field for an arbitrary charge distribution,''
  Proc.~R.~Irish Acad.~{\bf 81} (1947) 191.

%\cite{Lemos:2005md}
\bibitem{Lemos:2005md}
  J.~P.~S.~Lemos and V.~T.~Zanchin,
  ``A class of exact solutions of Einstein's field equations in higher-dimensional spacetimes, $d\ge 4$: Majumdar-Papapetrou solutions,''
  Phys.\ Rev.\ D {\bf 71} (2005) 124021
  [gr-qc/0505142].
  %%CITATION = GR-QC/0505142;%%

%\cite{Israel:1972vx}
\bibitem{Israel:1972vx}
  W.~Israel and G.~A.~Wilson,
  ``A class of stationary electromagnetic vacuum fields,''
  J.\ Math.\ Phys.\  {\bf 13} (1972) 865.
  %%CITATION = JMAPA,13,865;%%

%\cite{Perjes:1971gv}
\bibitem{Perjes:1971gv}
  Z.~Perj\'es,
  ``Solutions of the coupled Einstein Maxwell equations representing the fields of spinning sources,''
  Phys.\ Rev.\ Lett.\  {\bf 27} (1971) 1668.
  %%CITATION = PRLTA,27,1668;%%

%\cite{Hartle:1972ya}
\bibitem{Hartle:1972ya}
  J.~B.~Hartle and S.~W.~Hawking,
  ``Solutions of the Einstein-Maxwell equations with many black holes,''
  Commun.\ Math.\ Phys.\  {\bf 26} (1972) 87.
  %%CITATION = CMPHA,26,87;%%

%\cite{Gibbons:1982fy}
\bibitem{Gibbons:1982fy}
  G.~W.~Gibbons and C.~M.~Hull,
  ``A Bogomol'nyi bound for general relativity and solitons in $N=2$ supergravity,''
  Phys.\ Lett.\ B {\bf 109} (1982) 190.
  %%CITATION = PHLTA,B109,190;%%

%\cite{Tod:1983pm}
\bibitem{Tod:1983pm}
  K.~P.~Tod,
  ``All metrics admitting supercovariantly constant spinors,''
  Phys.\ Lett.\ B {\bf 121} (1983) 241.
  %%CITATION = PHLTA,B121,241;%%

%\cite{Bena:2009en}
\bibitem{Bena:2009en}
  I.~Bena, S.~Giusto, C.~Ruef and N.~P.~Warner,
  ``Multi-center non-BPS black holes: the solution,''
  JHEP {\bf 0911} (2009) 032
  [arXiv:0908.2121 [hep-th]].
  %%CITATION = ARXIV:0908.2121;%%

%\cite{Hartnoll:2009sz}
\bibitem{Hartnoll:2009sz}
  S.~A.~Hartnoll,
  ``Lectures on holographic methods for condensed matter physics,''
  Class.\ Quant.\ Grav.\  {\bf 26} (2009) 224002
  [arXiv:0903.3246 [hep-th]].
  %%CITATION = ARXIV:0903.3246;%%

%\cite{Anninos:2013mfa}
\bibitem{Anninos:2013mfa}
  D.~Anninos, T.~Anous, F.~Denef and L.~Peeters,
  ``Holographic vitrification,''
  arXiv:1309.0146 [hep-th].
  %%CITATION = ARXIV:1309.0146;%%

%\cite{Kastor:1992nn}
\bibitem{Kastor:1992nn}
  D.~Kastor and J.~H.~Traschen,
  ``Cosmological multi - black hole solutions,''
  Phys.\ Rev.\ D {\bf 47} (1993) 5370
  [hep-th/9212035].
  %%CITATION = HEP-TH/9212035;%%

%\cite{Brill:1993tm}
\bibitem{Brill:1993tm}
  D.~R.~Brill, G.~T.~Horowitz, D.~Kastor and J.~H.~Traschen,
  ``Testing cosmic censorship with black hole collisions,''
  Phys.\ Rev.\ D {\bf 49} (1994) 840
  [gr-qc/9307014].
  %%CITATION = GR-QC/9307014;%%

%\cite{Klemm:2000vn}
\bibitem{Klemm:2000vn}
  D.~Klemm and W.~A.~Sabra,
  ``Charged rotating black holes in 5-D Einstein-Maxwell (A)dS gravity,''
  Phys.\ Lett.\ B {\bf 503} (2001) 147
  [hep-th/0010200].
  %%CITATION = HEP-TH/0010200;%%

%\cite{Klemm:2000gh}
\bibitem{Klemm:2000gh}
  D.~Klemm and W.~A.~Sabra,
  ``General (anti-)de~Sitter black holes in five dimensions,''
  JHEP {\bf 0102} (2001) 031
  [hep-th/0011016].
  %%CITATION = HEP-TH/0011016;%%

%\cite{Maeda:2009zi}
\bibitem{Maeda:2009zi}
  K.~-i.~Maeda, N.~Ohta and K.~Uzawa,
  ``Dynamics of intersecting brane systems -Classification and their applications-,''
  JHEP {\bf 0906} (2009) 051
  [arXiv:0903.5483 [hep-th]].
  %%CITATION = ARXIV:0903.5483;%%

%\cite{Maeda:2010ja}
\bibitem{Maeda:2010ja}
  K.~-i.~Maeda and M.~Nozawa,
  ``Black hole in the expanding universe with arbitrary power-law expansion,''
  Phys.\ Rev.\ D {\bf 81} (2010) 124038
  [arXiv:1003.2849 [gr-qc]].
  %%CITATION = ARXIV:1003.2849;%%

%\cite{Gibbons:2009dr}
\bibitem{Gibbons:2009dr}
  G.~W.~Gibbons and K.~-i.~Maeda,
  ``Black holes in an expanding universe,''
  Phys.\ Rev.\ Lett.\  {\bf 104} (2010) 131101
  [arXiv:0912.2809 [gr-qc]].
  %%CITATION = ARXIV:0912.2809;%%

%\cite{Chimento:2012mg}
\bibitem{Chimento:2012mg}
  S.~Chimento and D.~Klemm,
  ``Black holes in an expanding universe from fake supergravity,''
  JHEP {\bf 1304} (2013) 129
  [arXiv:1212.5494].

%\cite{Liu:2000ah}
\bibitem{Liu:2000ah}
  J.~T.~Liu and W.~A.~Sabra,
  ``Multicentered black holes in gauged D = 5 supergravity,''
  Phys.\ Lett.\ B {\bf 498} (2001) 123
  [hep-th/0010025].
  %%CITATION = HEP-TH/0010025;%%

%\cite{McVittie:1933zz}
\bibitem{McVittie:1933zz}
  G.~C.~McVittie,
  ``The mass-particle in an expanding universe,''
  Mon.\ Not.\ Roy.\ Astron.\ Soc.\  {\bf 93} (1933) 325.
  %%CITATION = MNRAA,93,325;%%

%\cite{Vaidya:1968zza}
\bibitem{Vaidya:1968zza}
  Y.~P.~Shah and P.~C.~Vaidya,
  ``Gravitational field of a charged particle embedded in a homogeneous universe,''
  Tensor (Japan) {\bf 19} (1968) 191.
  %%CITATION = TNSRA,19,191;%%

%\cite{Hayward:1997jp}
\bibitem{Hayward:1997jp}
  S.~A.~Hayward,
  ``Unified first law of black hole dynamics and relativistic thermodynamics,''
  Class.\ Quant.\ Grav.\  {\bf 15} (1998) 3147
  [gr-qc/9710089].
  %%CITATION = GR-QC/9710089;%%

%\cite{Vaidya:1966zzc}
\bibitem{Vaidya:1966zzc}
  Y.~P.~Shah and P.~C.~Vaidya
  ``The gravitational field of a charged particle embedded in an expanding universe,''
  Curr.\ Sci.\  {\bf 36} (1966) 120.
  %%CITATION = CUSCA,36,120;%%

%\cite{Hayward:1993wb}
\bibitem{Hayward:1993wb}
  S.~A.~Hayward,
  ``General laws of black hole dynamics,''
  Phys.\ Rev.\ D {\bf 49} (1994) 6467.
  %%CITATION = PHRVA,D49,6467;%%

%\cite{Hayward:2008jq}
\bibitem{Hayward:2008jq}
  S.~A.~Hayward, R.~Di Criscienzo, L.~Vanzo, M.~Nadalini and S.~Zerbini,
  ``Local Hawking temperature for dynamical black holes,''
  Class.\ Quant.\ Grav.\  {\bf 26} (2009) 062001
  [arXiv:0806.0014 [gr-qc]].
  %%CITATION = ARXIV:0806.0014;%%

%\cite{Patel:1999ej}
\bibitem{Patel:1999ej}
  L.~K.~Patel, R.~Tikekar and N.~Dadhich,
  ``Higher-dimensional analog of McVittie solution,''
  Grav.\ Cosmol.\  {\bf 6} (2000) 335
  [gr-qc/9909069].
  %%CITATION = GR-QC/9909069;%%

%\cite{Gao:2004gz}
\bibitem{Gao:2004gz}
  C.~J.~Gao and S.~N.~Zhang,
  ``Higher-dimensional Reissner-Nordstr\"om-FRW metric,''
  Gen.\ Rel.\ Grav.\  {\bf 38} (2006) 23
  [gr-qc/0411040].
  %%CITATION = GR-QC/0411040;%%




\end{thebibliography}
\end{document}